\documentclass[twoside,jaspp]{ametsoc}  

\usepackage{graphicx}

\lefthead{FALKOVICH, STEPANOV AND VUCELJA} \righthead{RAIN}
\received{September, 2003}
\sectionnumbers
\printfigures                   
\authoraddr{G. Falkovich, Physics of Complex Systems,
Weizmann Institute of Science, Rehovot 76100 Israel,
gregory.falkovich@weizmann.ac.il}     

\begin{document}
\bibliographystyle{ametsoc}

\title{Rain initiation time in turbulent warm clouds}

\author{Gregory Falkovich}
\affil{Institute for Advanced Study, Princeton and Weizmann
Institute of Science, Israel}
\author{Mikhail G. Stepanov}
\affil{Institute for Advanced Study, Princeton and Weizmann
Institute of Science, Israel}
\author{Marija Vucelja}
\affil{Belgrade University, Yugoslavia and Weizmann Institute of
Science, Israel}

\begin{abstract}
We present a mean-field model that describes droplet growth due to
condensation and collisions and droplet loss due to fallout. The
model allows for an effective numerical simulation. We study how
 the rain initiation time depends on different parameters. We
also present a simple model that allows one to estimate the rain
initiation time for turbulent clouds with an inhomogeneous
concentration of cloud condensation nuclei. In particular, we show
that over-seeding even a part of a cloud by small hygroscopic
nuclei one can substantially delay the onset of precipitation.

\end{abstract}

\begin{article}

\section{Introduction}

Droplets start growing by vapor condensation on cloud condensation
nuclei (CCN) which are typically submicron-size particles. In warm
clouds,  coalescence due to collisions contributes the growth
until the raindrops (generally exceeding millimeters) fall out of
the cloud, see e.g. Pruppacher and Klett (1997). Those processes can
be modelled by the equation for the local distribution of droplets
over sizes, $n(a,t,{\bf r})=n(a)$ and the mass content of the
water vapor, $M(t,{\bf r})$:
\begin{eqnarray}
&&\!\!\!  {\partial n(a)\over \partial t}+({\bf v}\cdot\nabla)n =
 -{\kappa sM\over\rho_0} {\partial \over\partial a}{n(a)\over a}
 \label{Smo1}\\
 &&\!\!\!\!\!\! \!\!\! + \int da' \left[ \frac{K(a',a'')n(a')n(a'')}
 {2(a''/a)^2} - K(a',a) n(a')n(a)
\right]\,, \nonumber\\&& \!\!\!\!\!\!\!\!\!  {\partial M\over
\partial t}+({\bf v}\cdot\nabla)
M-\kappa\Delta M=-4\pi s\rho_0\kappa \int a n(a)da\ .\label{dens}
\end{eqnarray}
Here $a$ is the droplet radius, $t$ is time and ${\bf r}$ is the
coordinate in space. The first term in the right-hand side of
(\ref{Smo1}) is due to condensation which (for not very large
droplets) changes the droplet size $a$ according to
${da^2/dt}=\kappa sM/\rho_0$ where $\kappa$ is vapor diffusivity,
$s$ is the degree of supersaturation and $\rho_0=10^3\,{\rm
kg}\cdot{\rm m}^{-3}$ is the density of liquid water. The time
needed for condensation to grow a droplet of size $a$ is generally
proportional to $a^2$. The second term in the rhs of (\ref{Smo1})
describes coalescence due to collisions, here
$a''=(a^3\!-\!a'^3)^{1/3}$ is the size of the droplet that
produces the droplet of size $a$ upon coalescence with the droplet
of size $a'$. The collision kernel is the product of the target
area and the relative velocity of droplets upon the contact:
$K(a,a')\!\simeq\!\pi (a\!+\!a')^2\Delta v$. According to the
recent precise measurements (Beard et al, 2002) the coalescence
efficiency of the droplets in the relevant intervals is likely to
be greater than $0.95$ we put it unity in our calculations.
Collisions change the concentration $n(a)$ on a timescale of order
$1/K(a,a_1)n(a_1)$ where collision kernel $K$ is a fast growing
function of droplet sizes.

Since condensation slows down and coalescence accelerates as the
size of droplets grow then one can introduce a crossover scale
$a_*$, determined by $K(a_*)n\simeq \kappa sM/\rho_0a_*^2$. The
growth up to the crossover scale is mainly due to condensation
while coalescence provides for the further growth. The crossover
scale $a_*$ depends on $n$ and is typically in tens of microns
(see below).

To describe the six unknown functions, $n,M,s,{\bf v}$, one must
also add the equation that describes the temperature change (that
determines $s$) and the Navier-Stokes equation for the velocity.
Such system cannot be possible solved numerically with any
meaningful resolution neither presently nor in a foreseeable
future. The main problem is a very complicated spatial structure
of the fields involved particularly due to cloud turbulence. Our
aim in this paper is to formulate some mean-field model which does
not contain spatial arguments at all. The requirements to this
model is that it must give the correct qualitative relations
between the parameters and reasonable quantitative description (at
least within the order of magnitude) of the real-world timescales.
According to the two basic phenomena involved (condensation and
collisions), the main problems in space-averaging the equations
are related to the proper description of the two phenomena: mixing
and diffusion of water vapor and the influence of cloud turbulence
on collisions. We address them in Sections~\ref{sec:seed} and
\ref{sec:num} respectively. We use the model to study the
evolution of $n(a,t)$ starting from sub-micron sizes all the way
to the moment when droplet fallout significantly decreases the
water content in the cloud. We shall call this moment the rain
initiation time and we study how that time depends on
initial vapor content and CCN concentration and on the level of
air turbulence.

\section{Growth by gravitational collisions}\label{sec:grav}

\begin{figure}
  \includegraphics[width=3in,angle=270]{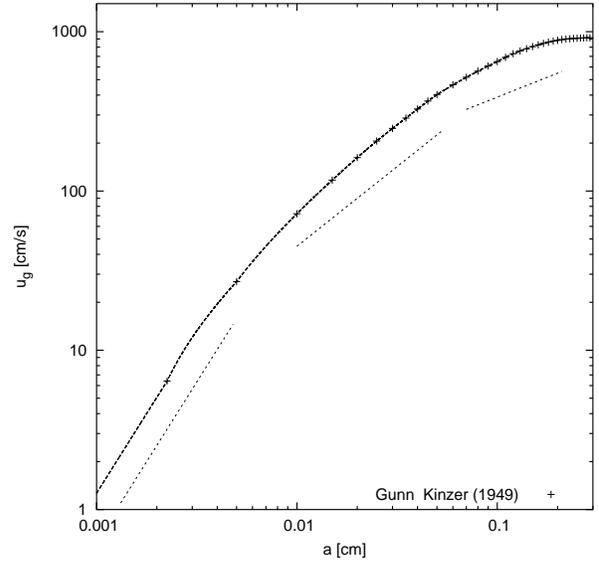}
  \caption{\label{fall}Terminal fall velocity.}
\end{figure}

For the parameters typical for warm precipitating clouds
($sM/\rho_0=10^{-8}\div 10^{-9}$ and $n=10^6\div10^9\,{\rm
m}^{-3}$), collisions are negligible for micron-size droplets
(Pruppacher and Klett 1997). For droplets larger than couple of
microns, Brownian motion can be neglected and the collision kernel
in a still air is due to gravitational settling:
\begin{equation}K_g(a,a')\!=\!\pi
(a\!+\!a')^2E(a,a')|u_g(a)\!-\!u_g(a')|\ .\label{Kg}\end{equation}
The fall velocity $u_g$ is obtained from the balance of gravity
force $4\pi g\rho_0a^3/3$ and the friction $F(u_g,a)$. The
friction force depends on the Reynolds number of the flow around
the droplet, $Re_a\!\equiv\! u_g a/ \nu$. When $Re_a$ is of order
unity or less, $F=6\pi\nu\rho au_g$ and $u_g\!=\!g\tau$ where
$\rho$ is the air density and
$\tau\!=\!(2/9)(\rho_0/\rho)(a^2/\nu)$ is called Stokes time. We
use $u_g=g\tau$ for $a<40\mu$m and take $u_g(a)$ from the
measurements of Gunn and Kinzer (1949) for $a>50\mu$m with a
smooth interpolation for $40\mu\,m<a<50\mu\,m$ as shown in
Fig~\ref{fall}. The dotted straight lines have slopes $2,1,1/2$.
One can see that $u_g\propto a^2$ at $a<40\,\mu{\rm m}$. There is
an intermediate interval with an approximately linear law
$u_g\propto a$ for $40\,\mu{\rm m}<a<400\mu{\rm m}$. When
$Re_a\gg1$ one may expect $F\propto \rho a^2 u_g^2$ as long as
droplet remains spherical; that gives $u_g\propto
\sqrt{ag\rho_0/\rho}$. Square-root law can be distinguished
between $400\,\mu{\rm m}$ and $1\,mm$ while the growth of $u_g(a)$
saturates at larger $a$ due to shape distortions. Hydrodynamic
interaction between approaching droplets is accounted in $K_g$ by
the collision efficiency $E$, which values we take from Pinsky et
al (2001) at the 750 mbar altitude.

It is of practical use to be able to predict the time left before
rain starts given the knowledge of droplet distribution at a given
instant. Such distributions can be measured  with high accuracy by
optical and other methods. Drop size distributions measured in
many different types of clouds under a variety of meteorological
conditions often exhibit a characteristic shape (Pruppacher and
Klett, 1997). Generally the concentration rises sharply from low
value maximum, and then decreases gently toward larger sizes,
causing the distribution to be positively skewed with a long tail
toward the larger sizes. We approximate such a shape with
half-Gaussian $\theta(a-a_0)\exp(-(a-a_0)^2/2\sigma^2)$ where
$\theta$ is a step function. We thus characterize the distribution
by two parameters: the mean position $P$ and the  width $\sigma$.
Since we mainly consider narrow initial distributions ($\sigma\ll
P$), the rain initiation time does not depend substantially on the
initial shape. We start from purely gravitational collisions in a
still air that is solve the space-homogenous version of
(\ref{Smo1}) with no condensation term:
\begin{eqnarray}
 && {\partial n(a)\over \partial t}=-n(a){u_g(a)\over
 L}\label{Smo2}\\&&\!\!\!\!\!\!\!\!\!+\!
 \int da' \biggl[ \frac{K_g(a',a'')n(a')n(a'')}{2(a''/a)^2}-
 K_g(a',a) n(a')n(a)
\biggr]\,.\nonumber
\end{eqnarray}
The first term in the rhs of (\ref{Smo2}) models the loss of
droplets falling with the settling velocity $u_g$ from the cloud
of the vertical size $L$. Since $L$ are generally very large (from
hundreds meters to kilometers) and $u_g(a)$ grows with $a$ (see
Fig.~\ref{fall} below), fallout is relevant only for sufficiently
large drops (called raindrops) with sizes millimeter or more. The
collision (Smoluchowsky) term describes the propagation of
distribution towards large sizes.  The asymptotic law of
propagation depends on the scaling of $K_g(a,a')$. If the
collision kernel is a homogeneous function of degree $\alpha$
[that is $K_g(\xi a,\xi a')=\xi^\alpha K_g(a,a')$] one can show
that for $\alpha$ larger/smaller than three the propagation is
accelerating/decelerating while for $\alpha=3$ it is exponential
$\ln a\propto t$ (see van Dongen and Ernst 1988; Zakharov et al
1991). Our numerics show, however, that the intervals of sizes $a$
where $\alpha$ is approximately a constant are too short for
definite self-similarity of the propagation to form both for
narrow and wide initial distributions. This is due to complexity
of both functions, $u_g(a)$ and $E(a,a')$. We thus focus on the
most salient feature of the propagation, namely study how the
amount of water left in the cloud, $W$, depends on time. The
decrease of that amount is due to a concerted action of collisions
producing large drops and fallout.

The droplets radii space was discretized, i.e. the droplets size
distribution $n(a, t)$ was presented as the set of concentrations
$n_i(t)$ of droplets with radius $a_i$.
The total mass of vapor and water in droplets is conserved in our calculations.
The grid of radii was
taken approximately exponential at sizes that are much larger than the size of initial
condensation nuclei, with 256 points in unit interval of natural
logarithm. The collision term in Smoluchowsky equation was treated
as follows: let the radius $(a_i^3 + a_j^3)^{1/3}$ of the droplet
resulted from merging of the two with radii $a_i$ and $a_j$ to be
in between of two radii $a_k$ and $a_{k+1}$ from the grid.  Then
the collision results in decreasing of $n_i$ and $n_j$ by quantity
$dN$ that is determined by the collision kernel, while the
concentrations $n_k$ and $n_{k+1}$ are increased in such a way
that sum of their change is $dN$ and the whole amount of water in
droplets is conserved in coalescence:
\begin{eqnarray}
&& \delta n_i = \delta n_j = -dN=-\delta n_k - \delta n_{k+1}\,,\nonumber\\
&&a_k^3 \delta n_k + a_{k+1}^3 \delta n_{k+1} = (a_i^3 + a_j^3) dN\,,\nonumber\\
&& \delta n_{k+1} = dN (a_i^3 + a_j^3 - a_k^3)/(a_{k+1}^3 - a_k^3)\,,\nonumber\\
&&\delta n_k = dN (a_{k+1}^3 - a_i^3 + a_j^3)/(a_{k+1}^3 - a_k^3)\ .
\label{col}\end{eqnarray}
The total
amount of water (the sum of the part that left and remained in the
cloud) was a conserved quantity, up to $10^{-6}$ accuracy, during
the whole simulation. Note that our scheme automatically keeps
the numbers positive: if $dN$ is greater than either $n_i$ or $n_j$,
then we choose $dN=\min\{n_i,n_j\}$, so that $n_i$ and $n_j$ are also not negative after
every elementary collision process. Let us stress that our scheme
is conservative both in mass and number of
droplets (comparing to the non-conservative scheme of
Berry and Reinhardt, 1974 and the scheme of Bott, 1998 which was
conservative only in mass and needed special choice of the time step
to keep positivity). The minimal time step needed for our calculations
was estimated from
characteristic timescales of our problem to be $0.1$ s. We have
checked that the decrease of the time step below $dt = 0.05~{\rm
s}$ does not change the results, the figures below all correspond
to that $dt$.

\begin{figure}
  \includegraphics[width=2.3in,angle=270]{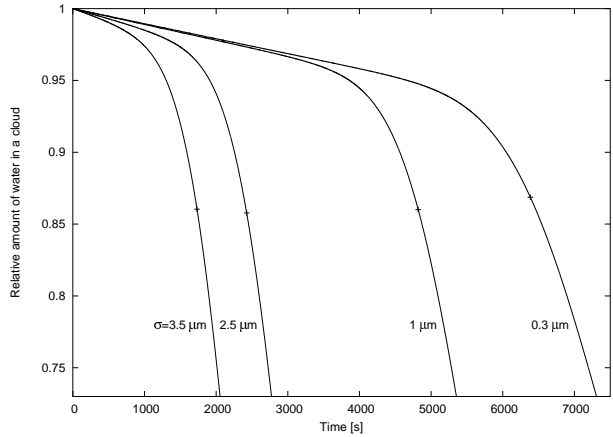}
  \caption{\label{water13}Fraction of water left in the cloud
  as a function of time.
  $P=13~\mu{\rm m}$.}
\end{figure}

\begin{figure}
  \includegraphics[width=2.3in,angle=270]{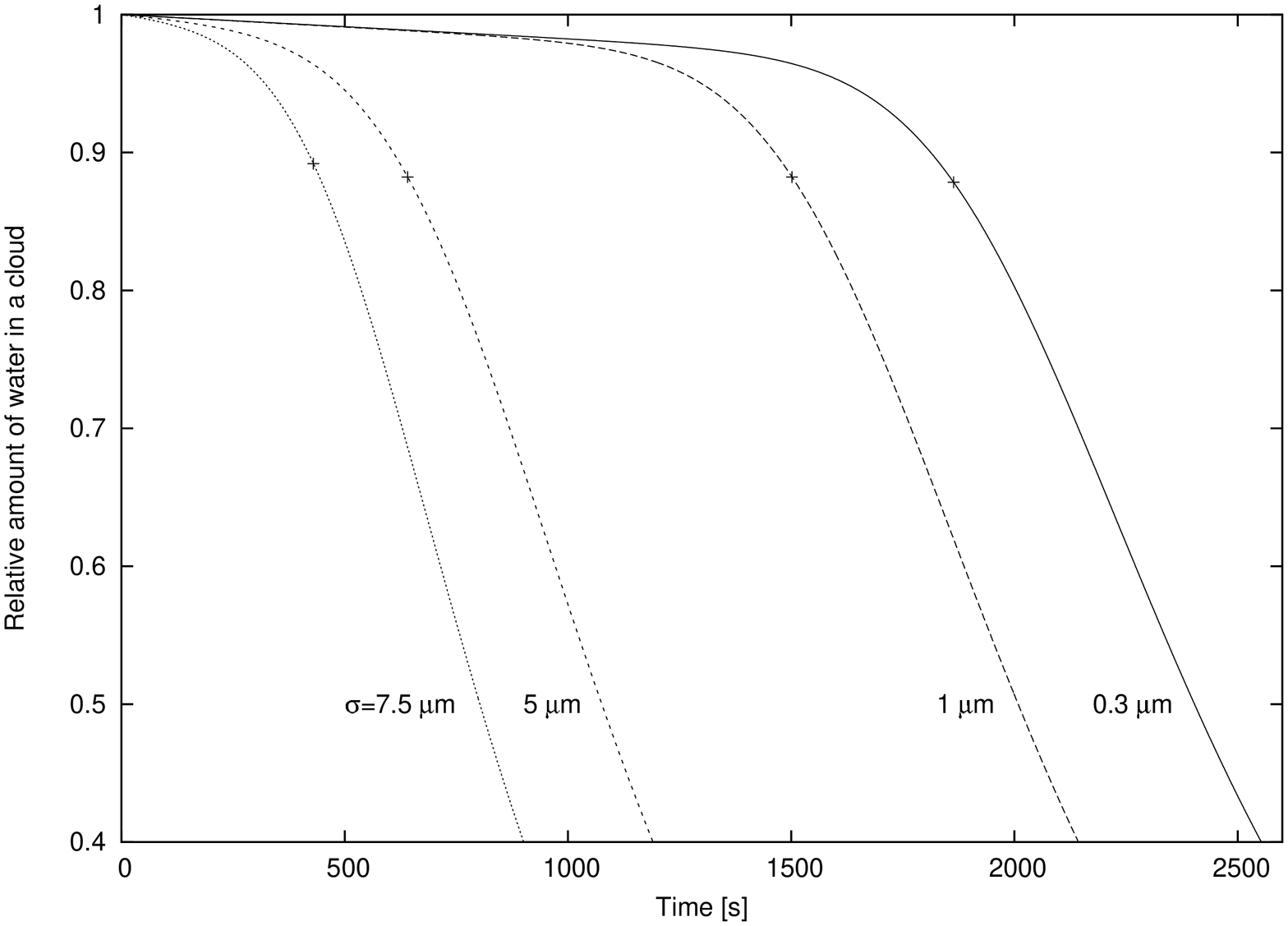}
  \caption{\label{water16} Fraction of water left in the cloud
  as a function of time.
  $P = 16~\mu{\rm m}$.}
\end{figure}

\begin{figure}
  \includegraphics[width=2in,angle=270]{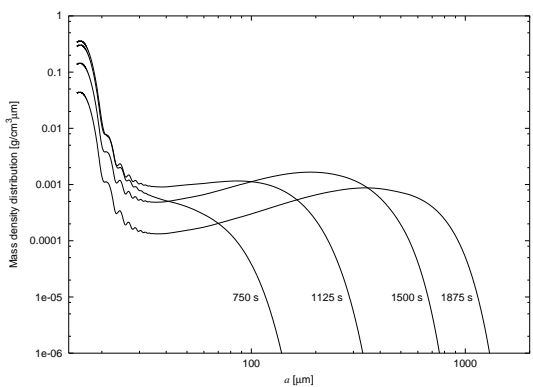}
  \caption{\label{mass}Mass density of water, $t_* \simeq 1500~{\rm s}$.}
\end{figure}

The graphs $W(t)$ are shown at Figs~\ref{water13} and
\ref{water16} (for $L=2$ km) and they are qualitatively the same
both for narrow and wide initial distributions. At the initial
stage, $W$ decreases slowly due to the loss of drizzle. After
large raindrops appear, loss accelerates. At every curve, the star
marks the moment when respective $d^2W/dt^2$ are maximal. After
that moment, the cloud looses water fast so it is natural to take
$t_*$ as the beginning of rain. Figure~\ref{mass} shows how the
mass distribution over sizes, $m(a)\propto a^3n(a)$ evolves with
time. One can see the appearance of secondary peaks and
distribution propagating to large $a$. The moment $t_*$ seems to
correspond
 to
the highest value of the envelope of the curves $m(a,t)$ of the
coalescence-produced drops. One can see from Figure~\ref{mass}
that the peak at mass distribution is
around 200 microns and most of the droplets are below 500 microns  at $t=t_*$.
The same character of the evolution
$W(t)$ can be seen in the next section for the {\it ab
initio} simulations of (\ref{Smo1},\ref{dens}).

\begin{figure}
  \includegraphics[width=3in,angle=270]{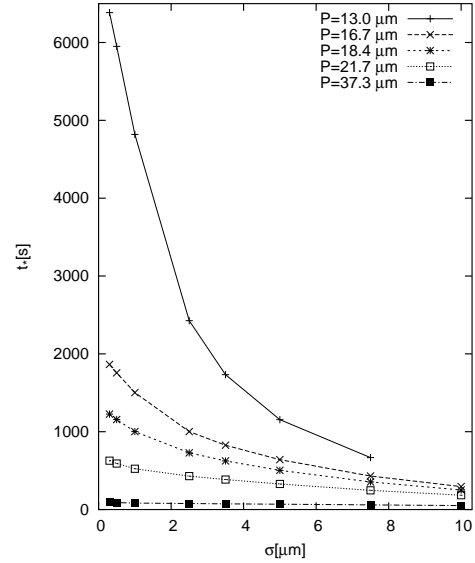}
  \caption{\label{RIT1} Rain initiation time as function
  of the width of initial distribution
  $\sigma$ for different initial positions $P$.}
\end{figure}

\begin{figure}
\includegraphics[width=2.3in,angle=270]{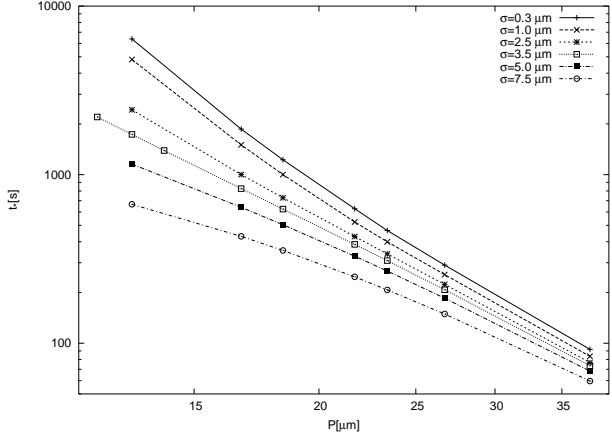}
\caption{\label{TP}Rain initiation time as function of the
position of the initial distribution.}
\end{figure}

\begin{figure}
\includegraphics[width=2.3in,angle=0]{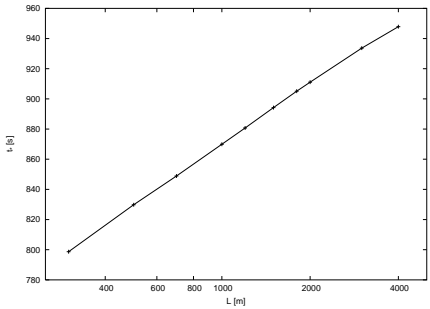}
\caption{\label{TL}Rain initiation time as function of the cloud
vertical size.}
\end{figure}

The rain initiation time $t_*$ defined in that way is presented in
Figures~\ref{RIT1} and \ref{TP} against the width and the position
of the initial distribution. Note the dramatic increase in $t_*$
with decreasing $\sigma$ for $P=13\,\mu$m. The mean droplet size
$P=14\,\mu$m is sometimes empirically introduced as the minimal
size required for the onset of precipitation (Rosenfeld and Gutman
1994). Figures~\ref{RIT1} and \ref{TP}  support that observation,
they indeed show that $t_*$ grows fast when P decreases below that
size but only for very narrow initial distributions and of course
there is no clear-cut threshold as $t_*(P)$ is a smooth (though
steep) function. The timescales (from tens of minutes to hours)
are in agreement with the data obtained before (see Pruppacher and
Klett, 1997, Chapter 15; and  Seinfeld, J. and S.Pandis, 1998,
Chapter 15 and the references therein). Figure~\ref{TP} also shows
that for $15\,\mu{\rm m}\lesssim P$, the function $t_*(P)$ can be
well-approximated by a power law $t_*\propto P^{-\gamma}$ with
$\gamma\approx 3$. The
 rain initiation time depends on the cloud vertical size
almost logarithmically as shown in Fig.~\ref{TL}, we do not have an
explanation for this functional form.

Here we treated the position and the width of the distribution as
given at the beginning of the collision stage. But of course the
distribution is itself a product of condensation stage so we now
turn to the consideration of the full condensation-collision
model.

\section{Condensation and collisions}\label{sec:num}

\begin{figure}\includegraphics[width=3in]{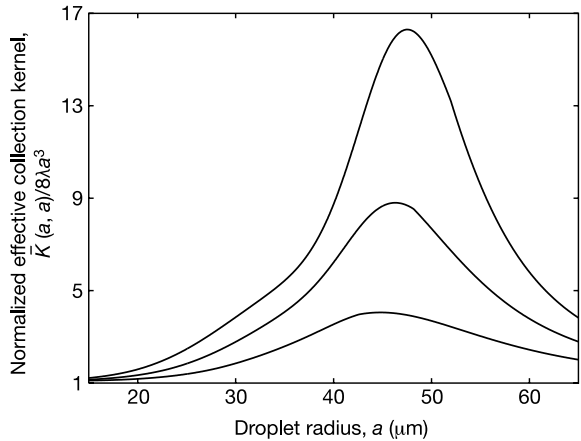}
\caption{\label{k12}Turbulence collision kernel normalized by $8\lambda a^3$
for equal-size droplets at ${\rm Re} = 10^6$.
From bottom to top, $\lambda = 10$, $15$ and $20~{\rm s}^{-1}$.}
\end{figure}

We consider now the space-homogeneous system
\begin{eqnarray}
&&  {\partial n(a)\over \partial t}=
 -{\kappa sM\over\rho_0} {\partial \over\partial a}{n(a)\over a} -
 n(a){u_g(a)\over L}
 \label{Smo3}\\
 &&\!\!\!\!\!\! \!\!\! + \int da' \left[ \frac{K(a',a'')n(a')n(a'')}
 {2(a''/a)^2} - K(a',a) n(a')n(a)
\right]\,, \nonumber\\&&  {\partial M\over
\partial t}=-4\pi s\rho_0\kappa \int a n(a)da\ .\label{dens1}
\end{eqnarray}
If one substitutes here gravitational and Brownian collision
kernels (taken, e.g. from Pruppacher and Klett 1997) and start
from $n=10^7\div10^8\,{\rm m}^{-3}$ sub-micron droplets growing in
a medium with $sM/\rho_0=10^{-8}\div 10^{-9}$ then
(\ref{Smo3},\ref{dens1}) give unrealistically large rain
initiation time. The reason for that is well-known: during the
condensation stage, the distribution shifts to larger sizes while
keeping its small width over $a^2$. For narrow distributions,
gravitational collisions are suppressed (since all droplets fall
with close velocities) as we have seen in the previous section.
Collisions of droplets with similar sizes are provided by an
inhomogeneous air flow. The velocity gradient $\lambda$ provides
for the kernel $K_s\!=\!\lambda (a\!+\!a')^3$ derived in Saffman
 and Turner (1956). However, typical velocity gradients in the air
turbulence ($\lambda\simeq 10-30\,{\rm s}^{-1}$) also do not
provide enough collisions (see e.g. Pruppacher and Klett 1997;
Jonas 1996; Vaillancourt and Yau 2000, and the referenced therein)
. Regular vertical inhomogeneity of supersaturation due to
temperature profile does not broaden $n(a)$ much even with the
account of turbulence-induced random fluctuations (Korolev 1995;
Turitsyn 2003). Spatial inhomogeneities in vapor content $M$ due
to mixing of humid and dry air still remains a controversial
subject (see. e.g Pruppacher and Klett 1997; Baker et al, 1980)
and probably can be neglected in cloud cores. We address the
turbulent mixing of vapor in Section~\ref{sec:seed} considering
partially seeded clouds. We address the turbulent mixing of vapour
in Section \ref{sec:seed} considering partially seeded clouds. As
far as collisions are concerned, the main effect of spatial
inhomogeneities seems to be the effect of preferential
concentration that is of turbulence-induced fluctuations in
droplets concentration (see Maxey 1987; Squires and Eaton 1991;
Sundaram and Collins 1997; Reade and Collins 2000; Grits et al
2000; Shaw et al 1998; Kostinski and Shaw 2001, and the references
therein). We use here the results of the recent theory by
Falkovich et al (2002). Namely, we multiply the Saffman-Turner
collision kernel $K_s$ by the enhancement factor $\langle
n_1n_2\rangle/\langle n_1\rangle\langle n_2\rangle$ (which
accounts for the effects of inertia and gravity in a turbulent
flow) and add the collision kernel due to the so-called sling
effect (droplets shot out of air vortices with too high
centrifugal acceleration), see Falkovich et al (2002) and
Falkovich and Pumir (2004) for the details. The total collision
kernel due to turbulence normalized by the homogeneous expression
$8\lambda a^3$ factor is presented in Fig.~\ref{k12} for
$Re=10^6$. We see that role of concentration inhomogeneities can
be substantial in the interval between 30 and 60 $\mu$m. The role
of sling effect is not significant at those levels of turbulence:
for the upper curve it gives the contribution of order of 10\%
between 25 and 35 $\mu$m.

The system (\ref{Smo3},\ref{dens1}) is our mean-field model where
the only memory of spatial inhomogeneities are the fallout term
and the renormalization of the collision kernel $K$. As we show
here, this model gives the rain initiation times with reasonable
quantitative values and proper qualitative behavior upon the change of
parameters. Let us discuss first how $t_*$ depends on $n$. Here, the
most important feature is the existence of the minimum in the function $t_*(n)$.
That can be explained by the competition between condensation and
collisions. Increasing $n$ one decreases $a_*$ (the crossover size
for which condensation time is comparable to the time of
collisional growth) and thus decreases the time needed for droplet
growth. This works until $a_*$ is getting comparable to $a_c\simeq
(M/n\rho_0)^{1/3}$. Indeed, when droplets grow comparable to $a_c$
vapor depletion slows and then stops condensation. If one takes
the initial concentration even larger so that $a_c<a_*$ then vapor
depletion stops condensation earlier and collisions are slower for
droplets of the smaller size $a_c$ so that the overall time of
droplet growth is getting larger. The concentration $n_*$ that
corresponds to the minimal time can be found from the (implicit)
relation $a_c\simeq a_*$ which corresponds to
\begin{equation}
(M/n_*\rho_0)^{-1/3}K\Bigl[(M/n_*\rho_0)^{1/3})\Bigr]\simeq\kappa
s\,\label{nstar}\end{equation}

That tells that $n_*\propto M$ and if $K\propto a^\alpha$ then
$n_*\propto s^{3/(1-\alpha)}$. One can argue that for small
concentrations (generally for maritime clouds), $n<n_*$, times of
condensation and collision stages are comparable. Therefore, $t_*$
is a function of the product $Ms$. For a homogeneous kernel,
$t_*\propto n^{-2/(2+\alpha)}(Ms)^{-\alpha/(2+\alpha)}$. For large
concentrations (generally for continental clouds), $n>n_*$, the
rain initiation time is mainly determined by collisions so it is
getting independent of the supersaturation and $t_*\propto
n^{(\alpha-3)/3} M^{-\alpha/3}$.

\begin{figure}
  \includegraphics[width=3in,angle=0]{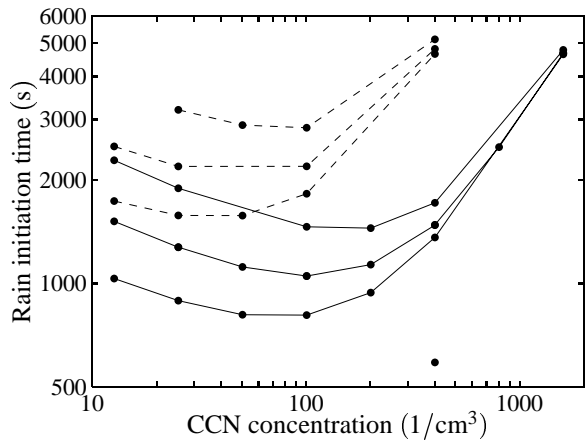}
  \caption{\label{ccn} Rain initiation time as function of CCN concentration
  $n$ for different supersaturations $s$ and  vapor contents $M$.}
\end{figure}

By numerically solving (\ref{Smo3},\ref{dens1}) with both gravity
and turbulence-induced collisions we obtain the rain initiation
time (also defined by the maximum of $d^2W/dt^2$) as a function of
the CCN concentration $n$ for different values of the
supersaturation $s$ and the vapor content $M$. The grid of radii was
approximately exponential at sizes that are much larger than the size of initial
condensation nuclei (with 200 points in unit interval of natural
logarithm).  The condensation of vapor was taken into account by
working on evolving grid of radii $a_i(t)$ keeping conserved the total mass
of water in droplets and vapor. Collisions were modelled according to
(\ref{col}) described above. Note that the numerical scheme we employ here
has an additional advantage (comparing to those described in
Pruppacher and Klett, 1997;
Berry and Reinhardt, 1974; Bott, 1998) of accounting simultaneously for
condensation and collisions while respecting conservation laws.
We used the time step
$dt= 0.01~{\rm s}$ during the condensation phase, on a later stage (dominated
by coalescence) $dt= 0.1~{\rm s}$ was enough.
Those results are
presented in Figure~\ref{ccn} for $L=2$ km and $\lambda=20$ s$^{-1}$.
The solitary point at the
lower part corresponds to $M = 6$ g$\cdot$ m$^{-3}$, $s=1/150$.
The three solid lines correspond to $M = 3$ g$\cdot$ m$^{-3}$
while the three dashed lines to $M = 1.5$ g$\cdot$ m$^{-3}$.
Inside the triplets, the lines differ by the values of the
supersaturation, from bottom to top, $s = 1/75,\,1/150,\, 1/300$.
 We see that indeed the graphs $t_*(n)$ all have
minima. The position of the minimum is proportional to $M$ as
expected and approximately proportional to $s^{-1/2}$ which would
correspond to $\alpha\simeq 7$ in this interval of sizes. We see
that the left parts of different curves with the same product $sM$
approach each other as $n$ decreases. To the right of the minima,
the curves with different $s$ but the same $M$ approach each other
as $n$ increases. That supports the previous conclusions on the
respective roles of condensation and collisions in determining the
rain initiation time.

\section{Delaying rain by hygroscopic
over-seeding}\label{sec:seed}
That the rain time is a non-monotonic function of the concentration of
droplets may provide a partial explanation for the conflicting
observations of the effect of hygroscopic seeding. By seeding
clouds with hygroscopic aerosol particles one can vary the number
of cloud condensation nuclei and thus the number of small droplets
at the beginning of the cloud formation. It was observed that such
seeding in some cases suppresses precipitation (see e.g. Rosenfeld
et al 2001), while in other cases enhances and accelerates it
(Cotton and Pielke, 1995; Mather 1991), see also Bruintjes (1999)
for a recent review.

It is often desirable to postpone rain, for instance, to bring
precipitation inland from the sea. The fact that $t_*$ grows when
$n > n_*$ suggests the idea of over-seeding to delay rain. This is
considered to be unpractical: ``It would be necessary to treat all
portions of a target cloud because, once precipitation appeared
anywhere in it, the raindrops \ldots would be circulated
throughout the cloud \ldots by turbulence'' (Dennis, 1980). We
think that this conclusion ignores another, positive, aspect of
cloud turbulence namely the mixing and homogenization of partially
seeded cloud during the condensation stage. Let us describe
briefly how it works for two cases.

\begin{figure}
  \includegraphics[width=3in]{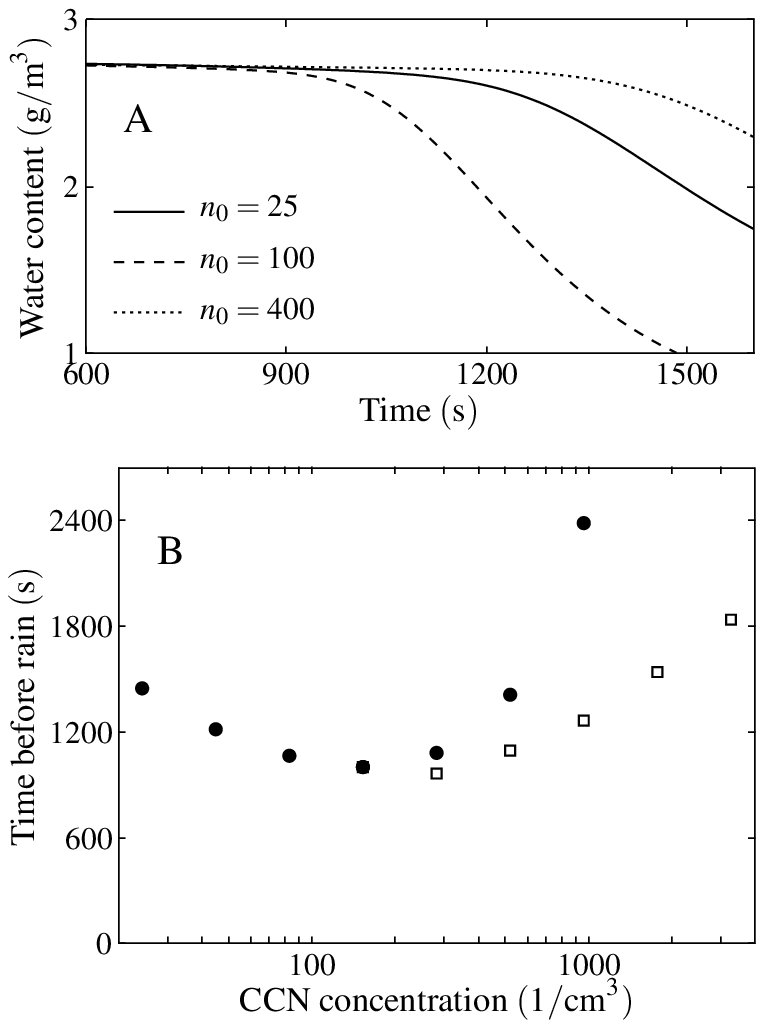}
  \caption{\label{full} Fraction of water left in the cloud as a
  function of time (A).
  Time of rain $t_*$ as a function of CCN concentration $n_0$ (B).
  The lower
    part (boxes) corresponds to a half-seeded cloud
    (the half-sum of concentrations is used as abscissa).}
\end{figure}

Consider first seeding a part of the cloud comparable to its size
$L_c$. Note that we do not consider here adding ultra-giant
nuclei, we assume seeded CCN to be comparable in size to those
naturally present. According to the Richardson law, the squared
distance between two fluid parcels grows as $\epsilon t^3$ so that
the rms difference of vapor concentrations between seeded and
unseeded parts decreases as $t^{-9/4}$ when $t^3 > t_0^3 =
L^2_c/\epsilon$ ($\epsilon$ is the energy dissipation rate in
turbulence). To see how different rates of condensation interplay
with turbulent mixing we generalize the mean-field system
(\ref{Smo3},\ref{dens1}) describing seeded and unseeded parts by
their respective $n_1,n_2$ and $x_1=s_1M_1,\,x_2=s_2M_2$ and link
them by adding the term that models the decay of the difference:
$dx_i/dt = \ldots - (x_i - x_j) t (t + t_0)^{-2} (9/4)$. As a crude
model, we assume two parts to evolve separately until $t=2t_0$,
then we treat the cloud as well-mixed and allow for the collisions
between droplets from different parts. That actually
underestimates the effect of seeding and can be considered as
giving the lower bound for the time before rain. The results of
simulations are shown in Fig.~\ref{full} for $t_0 = 180~{\rm s}$,
$L=2$ km and $\lambda=20$ s$^{-1}$.
It is seen from  Fig.~\ref{full}A that the water content $W$ changes
similarly to what was shown in Figs.~\ref{water13}.\ref{water16}
and the rain initiation time is again determined by the maximum of $d^2W/dt^2$.
The respective times are shown against $n_0 = (n_1 + n_2)/2$ by
boxes in Fig.~\ref{full}B. The time
increase is less than for homogeneous seeding but is still
substantial. The fraction of the cloud still unmixed after the
time $t$ decreases by the Poisson law $\exp(-t/t_0)$. Taking
$n_1=100~{\rm cm}^{-3}$ one sees that for a time delay of $10~{\rm
min}$ one needs to seed by $n_2 \simeq 3000~{\rm cm}^{-3}$.

Second, consider seeding by $N$ particles a small part of the
cloud which (unseeded) had some $n_0$ and would rain after $t_*$.
After time $t_*$ the seeds spread into the area of size $(\epsilon
t_*^3)^{1/2}$ with the concentration inside the mixed region
decaying as $n(t_*) = N(\epsilon t_*^3)^{-3/2}$ (for stratiform
clouds one gets $N(\epsilon t^3)^{-1}$). To have an effect of
seeding, one needs $n(t_*) > n_0$ which requires $N
> 10^{15}$ for $n_0 = 50~{\rm cm}^{-3}$, $t_* = 10~{\rm min}$ and
$\epsilon = 10~{\rm cm}^2{\rm s}^{-1}$. With sub-micron particles
weighing $10^{-11}~{\rm g}$ that would mean hundreds of kilograms
which is still practical.

\section{Summary}
We believe that our main result is a simple mean-field model
(\ref{Smo3},\ref{dens1}) which demonstrates non-monotonic
dependence of the rain initiation time on CCN concentration. As
the CCN concentration increases, the rain initiation time first
decreases and then grows as shown in Figs.~\ref{ccn},\ref{full}.
The simple modification of this model for an inhomogeneous case
described in Sect.~\ref{sec:seed} shows that one can increase the
rain initiation time even for a cloud partially seeded by
hygroscopic aerosols.

We acknowledge support by the Ellentuck fund, by the Minerva and
Israel Science Foundations and by NSF under agreement No.
DMS-9729992. We are grateful to A. Khain, M. Pinsky and D.
Rosenfeld for useful discussions.


\bibliography{mybibfile}

\end{article}
\end{document}